\documentstyle[11pt,aaspp4]{article}
\received{21 January 1999}
\accepted{8 March 1999}
\slugcomment{Astrophysical Journal Letters, {\em in press.}}
\lefthead{Henry}
\righthead{Diffuse Background Radiation}
\begin{document}
\title{ Diffuse Background Radiation}
\author{Richard C. Henry}
\affil{Center for Astrophysical Sciences\\Henry A. Rowland Department of Physics
and Astronomy\\The Johns Hopkins University, Baltimore, MD 21218-2686; henry@jhu.edu}

\begin{abstract}
A new determination of the upper limit to the cosmic diffuse background radiation,
at $\sim$110 $nm$, of $300\, photons$ $s^{-1}\, cm^{-2}\, sr^{-1}\, nm^{-1}$,
is placed in the context of diffuse background measurements
across the entire electromagnetic spectrum, including new optical, infrared, visible, and $\gamma$-ray background
measurements.  The possibility that observed excess diffuse
{\em visible} radiation is due to redshifted cosmological Lyman~$\alpha$ recombination
radiation is explored.  Also, a new standard of units for the display of spectra is advocated.
\end{abstract}

\section{Introduction}

We have recently reported (Murthy, Hall, Earl, Henry, and Holberg 1998) a new, and sharply lower,
value for the upper limit to the background radiation from the universe
at $\sim$110 $nm$.  Here, I place our new measurement into the context of
diffuse background measurements that have been made at {\em all} frequencies from radio to
gamma ray, in this way bringing out the potential
significance of the new measurement.  I also gather other new diffuse
background measurements that have been reported recently --- in the microwave,
in the far infrared, in the visible, and in the $\gamma$-ray spectral regions.  The visible
background, in particular, may be directly related in its origin to a
mechanism that is strongly suggested by our new ultraviolet
background upper limit.

Display of the broad spectrum of the cosmic background radiation apparently began with
Longair and Sunyaev (1969), who unfortunately chose a display method (the plotting of log I$_{\nu}$, with $I_{\nu}$ 
expressed in
$ergs$ $ s^{-1}\, cm^{-2}\, sr^{-1}\, Hz^{-1}$) that exaggerates the importance, in terms of energy per
decade of frequency, of the radio background (compared with the
$\gamma$-ray background) by a factor of as much as $10^{17}$.  This inferior method of plotting was also used
by Henry (1991), and a similar 
plot appears as Figure 5.5 of Kolb
 and Turner (1990), which is taken from the comprehensive review by Ressell and Turner (1990).

If one is interested in {\em energy content,} the most meaningful units in which to display the spectrum of diffuse
radiation are, remarkably enough, $photons$ $s^{-1}\, cm^{-2}\, sr^{-1}\, nm^{-1}$.  If
there is an {\em equal} amount of energy present in every logarithmic interval of
frequency, then these units have the virtue of assuming {\em constant} value --- as I now demonstrate:

For clarity, I omit ``$cm^{-2}\, s^{-1}\, sr^{-1}$" from the units.  I use constants $ h=6.6261 \times 10^{-27}$ $erg$ $s$
and $c=2.9979 \times 10^{10} $ $cm$ $ s^{-1}$.  If we have N $photons$ $ nm^{-1}$, then in a 1 $nm$ passband we
have $Nhc/\lambda_{nm}\,10^{-7}$  ergs.  But

\[
\Delta \nu_{Hz} = \frac{-c}{ \lambda^2_{nm} 10^{-7}}\,\Delta \lambda_{nm}
\]

So N $photons$ $ nm^{-1}$ corresponds to 

\[
 N\, photons\,nm^{-1}
 =Nh\lambda_{nm}\,ergs\, Hz^{-1}=\frac{Nhc}{\nu 10^{-7}} \,ergs\, Hz^{-1}\equiv \,I_\nu \,ergs\, Hz^{-1}
\]

\[
=N\, 1.9864\times 10^{-9}\, \frac{ergs\, Hz^{-1}}{\nu}
\simeq \frac {N}{5 \times 10^8} \frac{ergs\, Hz^{-1}}{\nu}
\]

If N is independent of frequency (a flat spectrum) then

\[
\int_{\nu '}^{b\nu '}{N\,photons\,nm^{-1}\,d\nu}=\int_{\nu '}^{b\nu '}{ \frac {Nhc}{\nu\, 10^{-7}}\,ergs\,Hz^{-1} d\nu}
=\frac{Nhc}{10^{-7}}\,ln(b) \: ergs
\]

This demonstrates that, whatever the frequency from which we integrate, as long as we
integrate over a specified factor $b$ in frequency we will obtain the same
amount of energy.  Note that it does not matter what you plot your value {\em against} --- constant
is constant.  It would be most consistent to plot your values against the natural
logarithm of the frequency; however, in Fig. 1 logarithms to base 10 are used.  If you wish the {\em area of paper} on your graph to be
 proportional to energy, you must create a plot that is {\em linear} in the proposed units, against the
 logarithm of the frequency (or energy).

It was recognized during the 1960's that some truer method than the plotting of $I_{\nu}$ is needed for the
 display of spectra. The method adopted, however, was {\em not} the use of the presently advocated units ; instead,
it was in effect reasoned that, if the spectrum is N~$photons$ $nm^{-1}$, then
$I_{\nu}=Nhc/\nu 10^{-7}$ $ ergs$ $ Hz^{-1}$, and if N is independent of $\nu$ (meaning, as
we see above, constant energy per
decade), then $\nu I_{\nu}=Nhc/10^{-7}$ $ergs$ is independent of $\nu$: so, one should plot {\em that}
, because it is {\em flat!}  Gehrels (1997) points to many recent references that employ such plots, showing that
use of plots of  $\nu I_{\nu}$ may be on its way to becoming an unfortunate new standard.  
Clearly, instead, one should simply plot N~$photons$ $nm^{-1}$
 (or, in full,  N $photons\, s^{-1}\, cm^{-2}\, sr^{-1} \,nm^{-1}$)
which I do, in Figure 1, for the background radiation spectrum of the universe.  

This plotting method has 
the great advantage that what is plotted is the detected ``quantity", ``per passband," that is, it is a {\em spectrum}; whereas
$\nu\,I_\nu$ is simply energy.   Note that the first and last parts of the last equation can be written

\[
\int_{\nu '}^{b\nu '}{N\,photons\,nm^{-1}\,d\nu}
=\nu I_{\nu}\times ln(b) \: ergs
\]

and note the presence, on the right hand side, of the factor $ln(b)$.  It has been emphasized by Gehrels (1997) that
{\em none} of those who use the increasingly ubiquitous plots of $\nu I_{\nu}$ (versus, ``whatever") include that factor,
which is therefore unity by implication, and so the ``constant quantity" that is plotted
 is energy {\em per natural logarithmic frequency,} again 
by implication.  Gehrels points out that frequently authors incorrectly state in such papers that what is plotted is energy
per decade, or energy per octave.  Gehrels also points out that integration of $\nu I_{\nu}$ is tricky, which is hardly
surprising, considering that it is an already-integrated quantity itself. Now, {\em none} of these 
considerations is present, if instead what is plotted, is what I have {\em advocated}
be plotted,  namely, the integrand in the first part of the last equation.  In view of its many
defects, the use of $\nu I_{\nu}$ should be permanently abandoned.

None of what I say should be taken to suggest that it is not {\em sometimes} appropriate to 
plot diffuse background (or other) spectra in other units.
Mather et al. (1990) display the 2.7 K background spectrum in units that, entirely appropriately,
exaggerate the highest energy part of the spectrum, which is where they made their brilliant measurements.  
Figure 6.2 of Peebles (1993) offers a complementary example of constructive display.

\section{The Background Spectrum}

I have assembled the background radiation spectrum of the universe in Figure 1.  The various contributions 
are discussed below.

\subsection{Radio Background}

The radio background spectra shown are from the Galactic Pole and the Galactic Plane, from Allen (1973).  The 
curvature of the spectra is due (Yates and Wielebinski 1967) to free-free absorption of the synchrotron
radiation by the partially-ionized Galactic disk.

\subsection{Microwave Background} 

The microwave background is shown for a temperature of 2.714K (Fixen et al. 1994).

\subsection{FIRAS Excess}

The spectrum (Fixsen et al. 1998), with $\pm 1 \sigma$ error range,  of the extragalactic microwave background in excess of the 2.714K black body 
background is shown. This represents a major discovery, comprising as it does about 20\% of the 
total intensity expected from the energy release from nucleosynthesis throughout the history of the universe.

\subsection{Infrared (DIRBE)}

I present the infrared background (detections at $140 \mu$ and $240 \mu$ are shown with error bars; the other points
are all upper limits) of Hauser et al. (1998).  The two highest upper limits are at $25\mu$ and $60\mu$, where
the interplanetary dust is brightest.  As just mentioned, these positive detections represent a major discovery for
cosmology.  The fact that these DIRBE detections, and the FIRAS detections previously mentioned, are in agreement,
is of course very satisfactory.

\subsection{Optical Background}

The optical background as evaluated by Bernstein (1998) is shown as solid circles with error bars.  Bernstein
points out the the level she finds is ``at least a factor 2 or 3" above the Hubble Deep Field integrated brightness
of galaxies, which is shown in Figure 1 as the line below Bernstein's points.  She suggests undetected outer 
envelopes of galaxies as a possible explanation for the excess radiation, but this is ruled out 
by the important finding of Vogeley (1997) that
the background of the Hubble Deep Field is smooth and cannot be made up of the integrated light of fainter galaxies.

The thin solid line through Bernstein's observational points is the extrapolation of the ultraviolet background
radiation to longer wavelengths, as we discuss next.

\subsection{Ultraviolet Background}

The ultraviolet background radiation has been reviewed by Henry (1991), and a figure giving the detailed spectral
observations appears in Henry and Murthy (1994).  The new Voyager upper limit of Murthy et al., of 
 $300\, photons$ $s^{-1}\, cm^{-2}\, sr^{-1}\, nm^{-1}$, is shown in our 
Figure 1 as a solid triangle, joined to the ultraviolet observations longward of Lyman $\alpha$ by a vertical
line at 121.6 nm.  The ultraviolet observations longward of Lyman $\alpha$ (Henry and Murthy 1994) 
are summarized here simply by a solid line.  That line
is the model of Henry and Murthy (1994), which is the spectrum of redshifted Lyman $\alpha$ recombination radiation
from ionized intergalactic clouds.  These clouds must be substantially clumped, and their ionization must be 
maintained by unknown means (although the suggestion by Sciama (1997) that neutrinos decay with the emission of an
ionizing photon would do the job.)  The horizontal ``error bar" at $\sim 10^{14}\,Hz$ is identified by Gnedin and
Ostriker (1997) as the redshifted Lyman $\alpha$ frequency interval ($z$ = 10 to 20) where maxiumum Lyman $\alpha$
emission is expected due to the re-heating (leading to the re-ionization) of the universe.

\subsection{Extreme Ultraviolet:  Optical Depth}

In the extreme ultraviolet the interstellar medium is very opaque, and instead of showing the observed background, 
which is entirely local, I choose to show (right hand scale in Figure 1) the logarithm of the
 photoionization 
optical depth (for hydrogen columns  
of $10^{19}\,cm^{-2}$, $10^{18}\,cm^{-2}$, and $10^{17}\,cm^{-2}$), obtained using the cross-sections of Morrison and 
McCammon (1983).

\subsection{Soft X-Ray Background}

The soft X-ray background has been reviewed by McCammon and Sanders (1990); their reports of the measurements are shown
as the seven small boxes that are at lowest intensities.  The width 
of each box has no meaning, while  the height of each box is the {\em range} of observed values, from low to high Galactic
latitudes (excluding special regions).  

Above these seven boxes are plotted the earliest observations of the soft X-ray background.  The highest
box is that of Bowyer, Field, and Mack (1968) which shows their extrapolation to extragalactic intensity, which, they stated,
may reasonably be interpreted as a continuation of the background spectrum already observed above 1 keV.  Below their
box is the observed intensity of Henry et al. (1968), which they correctly identified as a new component of X-ray emission,
but which they incorrectly attributed to emission from intergalactic
gas.  At slightly higher energy is the observation (very small filled box) of Henry et al. (1971): the emission was again attributed to
emission from intergalactic gas, this time, perhaps, correctly (Wang and McCray 1993).  The box contiguous below is
the confirming observation of Davidsen et al. (1972), which is in reasonable agreement with that reported by McCammon and Sanders.

\subsection{High Energy Background}

From log $\nu$ = 18, I plot the X-ray background spectrum of Boldt (1987), in addition extrapolating his spectrum to longer wavelengths 
(dashed line) to make clear how  {\em extraordinary} is the excess, that was first recognized by Henry et al. (1968), that occurs in the 
soft X-ray region.  Superimposed on Boldt's
spectrum, and extending to higher energy, is the fit to the data of Gruber that is quoted by Fabian and Barcons (1992).  Finally,
above log $\nu=19.6$ I have plotted the high-energy background data and upper limits that are 
presented by Sreekumar et al. (1998); the famous ``MeV bump" (Fichtel et al. 1978) has now vanished.

\section{Discussion}

A virtue of having the entire background radiation spectrum of the universe presented in a single diagram, 
in units that allow comparison of relative energy content, is that the parts may be
seen in relation to the whole, and possible connections may be examined.  The X-ray background that was discovered by 
Giacconi et al. (1962), has, very slowly, been revealed
as largely due to the integrated radiation of faint point sources (Ueda et al. 1998).  The background 
below $10^{18}\,Hz$ is clearly of independent origin.  Henry et al. (1968) failed to focus
on the most important aspect of their observation, the detection of strong soft X-ray emission at {\em low} Galactic latitudes.
  That no such radiation existed was sufficiently strongly believed, at that time,  that Bowyer, Field, and Mack (1968) subtracted out all
  of their low-latitude signal as particle contamination.
  
The main focus of the present paper is the relation of the soft X-ray background to the ultraviolet and visible backgrounds.  Note
the very large jump in background intensity from $10^{17}\,Hz$ to $10^{15}\,Hz$.  Where exactly this jump occurs is unclear,
but the fact that our observed intensity between 91.2 nm and 121.6 nm is only an upper limit may be revealing.  The sharp jump
precisely at 121.6 nm is clearly extremely important if real, as we believe it to be.

\section{Conclusion}

I conclude that a new possible origin for the diffuse visible background has been identified, the extention of the observed ultraviolet
background.  A new upper limit to the background shortward of 121.6 nm reveals the ultraviolet (and visible) backgrounds to {\em perhaps}
be redshifted Lyman $\alpha$ recombination radiation from an ionized intergalactic medium.  If this is so, additional support is
provided for the idea of Sciama (1997) that much of the non-baryonic dark matter is massive neutrinos that decay with the emission
of an ionizing photon.  
While intriguing, none of these ideas can be accepted, yet, as facts; definitive measurement of the diffuse
ultraviolet background radiation spectrum is called for, before any secure conclusions can be adopted.

\acknowledgements
This work was supported by NASA grant NAG53251 to the Johns Hopkins University.

\clearpage

\figcaption{The background radiation spectrum of the universe --- 1.) Radio, 2.) Cosmic Microwave background,
3.) FIRAS excess (Fixen et al. 1998), 4.) DIRBE background (points with error bars) and DIRBE upper limits (Hauser et
al. 1998), 5.) Optical background (Bernstein 1998), 6) Ultraviolet background (Murthy et al. 1998, Henry and Murthy 1994, Henry 1991), 
7.) Interstellar Medium Photoionization optical depth (right hand scale) for $10^{19},10^{18}$, and $10^{17}$
H atoms $cm^{-2}$, 8.) Soft X-Ray background, and 9.) High Energy background.  In this diagram, equal
plotted value means equal amount of energy per logarithmic interval of frequency.  The new {\em Voyager}
upper limit of Murthy et al. (1998) between 912~\AA$\,$ and 1216~\AA$\,$ suggests that the transition
from the high background in the visible, to the low background in the X-Ray, may occur at 1216 \AA, which in 
turn would suggest that the ultraviolet and visible background at high Galactic latitudes is
redshifted Lyman $\alpha$ recombination radiation.\label{fig1}}

\end{document}